\documentclass[epj]{svjour}
\usepackage{graphics}

\usepackage{amsmath}
\begin{document}
\title{Shell model description of Gamow-Teller strengths in $pf$-shell nuclei}
\author{}
\author{Vikas Kumar$^1$\thanks{Electronic address: vikasphysicsiitr@gmail.com}, P. C.~Srivastava$^1$\thanks{Electronic address: pcsrifph@iitr.ac.in}}
\institute{Department of Physics, Indian Institute of Technology, Roorkee 247 667, 
India }
\date{\today}

\abstract{   
A systematic shell model description of the experimental Gamow-Teller transition strength distributions in $^{42}$Ti, $^{46}$Cr, $^{50}$Fe and $^{54}$Ni is presented.
These transitions have been recently  measured via $\beta$ decay of these $T_z$=-1 nuclei, produced in fragmentation reactions at GSI
and also with ($^3${He},$t$) charge-exchange (CE) reactions corresponding to $T_z = + 1$ to  $T_z = 0$ carried out at RCNP-Osaka.
 The calculations are performed in the $pf$ model space, using the GXPF1a and KB3G effective interactions.
Qualitative agreement is obtained for the individual transitions, while the calculated summed transition strengths closely reproduce  the observed ones.}


\authorrunning{Vikas Kumar and P.C. Srivastava}
\maketitle

\maketitle
\section{Introduction \label{intro}}

Beta decays and electron capture reactions play an important role in nuclear physics \cite{Fujita2011} and in
many astrophysical phenomena like supernovae explosions and nucleosynthesis \cite{Rolfs88,Langanke2003}.
In energetic contexts like supernova explosions neutrino capture reactions are also relevant \cite{Balasi2015}.

$\beta$ decay has a direct access to the absolute GT transition strengths B(GT), allowing the study of
half-lives, Q$_{\beta}$ -values and branching ratios in the Q-window. Charge exchange reactions  like
$(p,n)$  and ($^3$He,$t)$ are useful tools to study the relative values of B(GT) strengths up to high
excitation energies. Recent experimental improvements have made possible to make one-to-one comparisons
of GT transitions studied in charge exchange reactions and $\beta$ decays \cite{Fujita2011}. Employing
the isospin symmetry experimental information can be obtained for unstable nuclei.  A long series of
high quality experiments have provided new experimental information about the Gamow-Teller strength
distribution in medium mass nuclei employing these techniques \cite{Fujita2005,Fujita2014,Orrigo2014}.  

Theoretical investigation to study strong magnetic dipole ($M1$) transitions and $GT$ strengths for
$fp$ shell nuclei reported in refs. \cite{LZ1,LZ2,talmi,martinez1,richtler,kota,caurier,mar}.
The effects of $T=0$ two body matrix elements on $M1$ and Gamow-Teller transitions for $^{44}$Ti, $^{46}$Ti,
and $^{48}$Ti is reported in refs. \cite{sjq,zamick1}. In this work it was shown that transition rates were
much more sensitive to the details of the $T=0$ interaction. In another work Garcia and Zamick \cite{zamick2}
reported the effects of several different interactions on $B(GT)$ values and magnetic moments in a single $j$-shell
calculation. The magnetic dipole excitations of $^{50}$Cr \cite{pai} has been recently studied with the method of nuclear
resonance fluorescence up to 9.7 MeV using bremsstrahlung at Darmstadt. In this work it is shown that
spin $M1$ resonance is mainly generated by spin-flip transitions between the orbitals of the $fp$ shell.
In ref. \cite{petermann} the relation of low-lying $GT_{-}$ strength and $SU(4)$ symmetry reported. It was confirmed
that $SU(4)$ symmetry broken with increasing spin-orbit splitting and by increasing neutron numbers.
The understanding of $M1$ excitations for $fp$ shell nuclei is possible with recent theoretical and experimental efforts 
\cite{heyde}. The $GT$ strength at high excitations is also important, in ref. \cite{ber} it was shown that 50\% of the strength
is shifted into the region of 10-45 MeV excitation for the nucleus $^{90}$Zr.
In the recent experiment using RISING array the super-allowed Gamow -Teller decay of the doubly magic nucleus $^{100}$Sn reported in ref. \cite{nat}.
The experimental findings were interpreted using large scale shell model with valence space consists of the
fifth (4$\hbar w$) harmonic oscillator shell,
that is, proton and neutron $\pi\nu(g,d,s)$ orbitals outside the $^{80}$Zr core using  five particle–hole excitations
from the $g_{9/2}$ proton and neutron orbitals to the rest of the shell to get 
convergent results for excitation spectra and the Gamow - Teller strength.

Large-scale shell-model calculations, employing a slightly monopole-corrected version of the well-known
KB3 interaction, denoted as KB3G, were able to reproduce the measured Gamow-Teller strength distributions
and spectra of the $pf$ shell nuclei in the mass range A = 45-65 \cite{Caurier99}. The description of
electron capture reaction rates, and the strengths and energies of the Gamow-Teller transitions in
$^{56, 58, 60, 62, 64}$Ni required a new shell-model interaction, GXPF1J \cite{Suzuki11}.

Shell-model calculations in the $pf$ model space with the KB3G and GXPF1a interactions qualitatively
reproduced experimental Gamow-Teller strength distributions of 13 stable isotopes with 45$\leq$A$\leq$64.
They  were used to estimate electron-capture rates for astrophysical purposes with relatively good accuracy
\cite{Cole2012}.  Shell model diagonalizations have become the appropriate tool to calculate the allowed
contributions to neutrino-nucleus cross sections for supernova neutrinos \cite{Balasi2015}.

Recently, F. Molina {\it et. al.} \cite{Molina,Molinathesis}, populated the $^{42}$Ti, $^{46}$Cr, $^{50}$Fe and $^{54}$Ni nuclei by
the fragmentation of a $^{58}$Ni beam at 680 MeV/nucleon on a {400 mg/{cm}$^2$} {Be} target and studied the 
$\beta$-decay. With the help of experimentally observed $\beta$-decay half lives, excitation energies, and $\beta$ branching ratios, 
they reported the Fermi and Gamow-Teller transition strengths and compared them with the more precise 
B(GT) value reported in \cite{Adachi} with the help of charge-exchange reaction at high excitation energies,
finding very good agreement between the both 
experimental data.

The aim of the present study is to present state of the art shell model calculations
for the observed transitions in $^{42}$Ti, $^{46}$Cr, $^{50}$Fe and $^{54}$Ni nuclei, restricted to the $pf$ model space, employing the
KB3G \cite{Caurier99} and GXPF1a \cite{Honma2} interactions. The shell model calculations are performed using the code
NuShellX@MSU \cite{Brown}. They provide a theoretical description of the experimental results presented
in \cite{Molina,Molinathesis} and \cite{Fujita2005,Adachi}, 
complementing those presented in \cite{Fujita2014}. Thus present work will add more information to refs.
\cite{Fujita2014,Molina,Adachi}.


\section{Details about shell model calculations }

In order to describe the measured GT strength distribution for $^{42}$Ti, $^{46}$Cr, $^{50}$Fe, and $^{54}$Ni
nuclei we employ the shell-model restricted to the $pf$ valence space and the effective interactions KB3G and GXPF1a. 

The GXPF1a is based on the GXPF1 interaction. Honma et al. \cite{Honma1} derived the
effective interaction, GXPF1, starting from the Bonn-C potential, by modifying 70 well-determined combinations
of four single-particle energies and 195 two-body matrix elements by iterative
fitting calculations about 699 experimental energy data out of 87 stable nuclei. 

The GXPF1 interaction was tested extensively \cite{Honma2} performing shell model calculations in the full $fp$ shell 
for binding energies, electromagnetic moments and transitions, and excitation spectra in the wide range of
$fp$ shell nuclei. 
As the $N = 34$ subshell gap in Ca and Ti isotopes, predicted by the GXPF1 interaction,
was not observed in recent experimental studies of the $^{52-56}$Ti isotopes \cite{Honma3,Honma4}, 
it led to the modification of GXPF1 interaction. 
Five $T = 1$ two-body matrix elements in the $fp$ shell were modified: 3 pairing interaction matrix elements were made slightly weaker and
two quadrupole-quadrupole matrix elements were made slightly stronger \cite{Honma2}. The modified interaction, referred to
as GXPF1a \cite{Honma2}, gave an improved description simultaneously for all these three isotope chains and is considered reliable for the use
in shell model calculations to explain the data on unstable nuclei. 

The interaction KB3G \cite{Caurier99} is a monopole-corrected version of the previous KB3 interaction  in order to treat properly
the $N = Z = 28$ shell closure and its surroundings \cite{A.Poves}. The parameters were fitted using experimental energies
of the lower $fp$ shell nuclei.

In the present work we employ the interaction GXPF1a. It generates nearly indistinguishable
results with respect to GXPF1J, the one employed in \cite{Fujita2014}. Both seem to particularly
well suited to describe nuclei with $A \leq 50$. For lighter nuclei the interaction KB3G is able to
describe with more detail the low energy spectra. 

The full shell model Hilbert space in the $pf$ shell  is employed in the description of $^{42}$Ti, $^{46}$Cr and $^{50}$Fe nuclei. 
Due to the huge matrix dimensions, in the case of  $^{54}$Ni we allowed for a maximum  of four
nucleon excitation from the ${f_{7/2}}$ shell to the rest of the $pf$ orbitals.

\begin{figure}
\begin{center}
\resizebox{0.50\textwidth}{!}{
\includegraphics{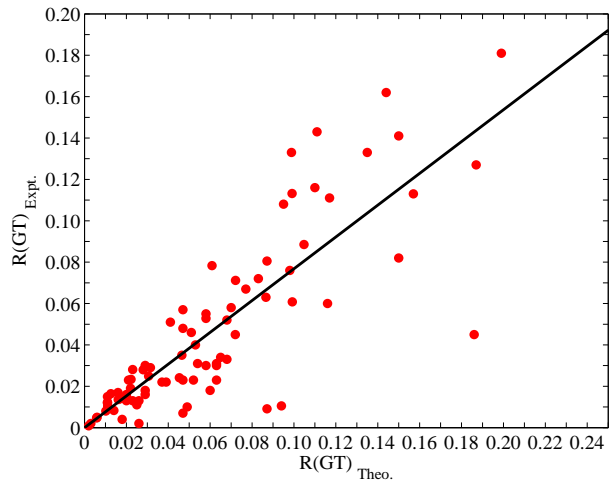} 
}
\end{center}
\caption{
Comparison of the experimental matrix elements R(GT) values with theoretical calculations based
on the ``free-nucleon" Gamow-Teller operator. Each transition is indicated by a point in the x-y plane. 
Theoretical and experimental value given by the x, y coordinates, respectively. }
\label{42Ti}
\end{figure}

\begin{table*}
 \begin{center}
    \leavevmode
    \caption{Experimental and theoretical $M(GT)$ matrix elements. The calculated shell model results with KB3G interaction.
    We have taken experimental data from ref. \cite{nndc}. Here
  $I_\beta + I_\epsilon$ are the branching ratios.}
    \label{tab:mgt_comp} 
    \begin{tabular}{lccccrrr}
    \hline
        &  & $Q$  & $I_\beta +
      I_\epsilon$ &  & \multicolumn{2}{c}{$M(GT)$} &  \\
      \cline{6-7}
     Process &$2J_n^\pi, 2T_n^\pi$ & (MeV) & (\%) & $\log ft$ & \multicolumn{1}{c}{Expt.} &
      \multicolumn{1}{c}{Theor.} & $W$ \\  
      \hline
      $^{43}$Sc($\beta^+){}^{43}$Ca & $7_1^-, 3$ & 2.221 & 77.54           & 5.04           &0.669  & 0.589 & 6.192 \\ 
                                    & $5_1^-, 3$ & 2.005 & 22.53            & 4.98          &0.720 & 0.682 &  \\ 
                                    & $5_2^-, 3$ & 0.290 & 0.025            & 5.68           &0.320 & 0.423 &  \\ 
      $^{44}$V($\beta^+){}^{44}$Ti & $4_1^+, 0$   & 12.341 & 32.0          & 4.70        &0.783  & 1.092 & 6.923 \\ 
                                   & $4_2^+, 0$   & 10.899 & 23.0          & 4.56        &0.920  & 0.941 &   \\ 
                                   & $4_3^+, 0$   & 9.315 &14.71          & 4.39        &1.119  & 0.995&  \\ 
      $^{44}$V($\beta^+){}^{44}$Ti & $12_1^+, 0$ & 9.415   & 56.01          & 4.10         & 2.519 & 2.361 & 11.163 \\ 
                                
      $^{45}$Ti($\beta^+){}^{45}$Sc & $7_1^-, 3$   & 2.062 &84.8              & 4.591         & 1.123 & 1.231 & 6.192\\
                                    & $7_2^-, 3$   & 0.654 &0.09             & 5.78         & 0.285 & 0.319 &    \\ 
      $^{46}$Sc($\beta^-){}^{46}$Ti & $8_1^+,2$    & 0.356 & 99.99          & 6.200         &0.187 & 0.222 & 13.136 \\ 
                                    
      $^{46}$V($\beta^+){}^{46}$Ti  & $2_1^+, 2$  & 2.735   &0.011              & 5.0       & 0.248 & 0.290 & 3.096  \\ 
                                    
     $^{47}$Ca($\beta^-){}^{47}$Sc   & $7_1^-, 5$   & 1.992 &27           & 8.3            & 0.016 & 0.028 & 16.383  \\
                                    & $5_1^-, 5$   &  0.695&73           & 6.08            & 0.202 & 0.177 &         \\
     $^{48}$Mn($\beta^+){}^{48}$Cr  & $8_1^+, 0$ & 11.642  & 6.50           & 5.4            & 0.469 & 0.381 & 9.288\\
                                    & $8_2^+, 0$ & 9.071 & 10.11           & 4.6            & 1.179 & 1.736 &       \\
     $^{49}$Ca($\beta^-){}^{49}$Sc  & $3_1^-, 7$  & 2.178 & 90.2          & 5.08        & 0.452 & 0.854 & 13.136 \\
                                    & $3_2^-, 7$  & 1.745 & 0.14         & 7.48        & 0.028 & 0.04 &  \\
                                    & $5_1^-, 7$  & 0.930 & 0.006         & 7.76        & 0.021 & 0.347 &  \\
                                    & $1_1^-, 7$  & 0.769 & 0.65         & 5.42        & 0.306 & 0.62 &  \\
     $^{49}$Sc($\beta^-){}^{49}$Ti  & $7_1^-, 5$   & 2.006 & 99.94           & 5.72         & 0.306 & 0.366 &  16.383\\  
                                    & $5_1^-, 5$   & 0.383 & 0.01           &7.0        & 0.070 & 0.299 &     \\       
     $^{49}$Cr($\beta^+){}^{49}$V  & $7_1^-, 3$  &2.627  & 12.43          & 5.60        & 0.304 & 0.251 & 5.363 \\
                                    & $5_1^-, 3$  & 2.536 &35.31         & 5.02        & 0.593 & 0.631 &      \\
                                    & $3_1^-, 3$  & 2.474 &50.54         & 4.81        & 0.756 & 0.806 &        \\
                                    & $5_2^-, 3$  & 1.113 &0.08         & 5.80        & 0.242 & 0.241 &        \\
                                    & $3_2^-, 3$  & 0.966 &0.03         & 6.15        & 0.161 & 0.153 &        \\
    $^{49}$Mn($\beta^+){}^{49}$Cr  & $7_1^-, 1$ & 7.443  & 5.81   & 4.80 & 0.764 & 0.594 & 5.363 \\            
    $^{50}$Sc($\beta^-){}^{50}$Ti  & $12_1^+, 6 $  & 3.691 &   0.58     & 7.01           & 0.269 & 0.467 & 20.538 \\
                                   & $10_1^+, 6 $  & 2.009 & 0.21            & 6.37            & 0.170 & 0.281&         \\
  $^{50}$Mn($\beta^+){}^{50}$Cr  & $2_1^+, 2 $  & 4.006 &   0.06   & 5.14           & 0.211 & 0.284 & 5.363 \\
                                 & $2_2^+, 2 $  & 2.636 &0.0007            & 5.90            & 0.088 & 0.068&   \\
  $^{50}$Mn($\beta^+){}^{50}$Cr  & $12_1^+, 2 $  & 4.470 & 8.05   & 6.0           & 0.260 & 0.248 & 11.163 \\
                                 & $8_1^+, 2 $  & 4.309 &69.46            & 5.0            & 0.822 & 0.974& 9.289\\
                                 & $12_2^+, 2 $  & 3.809 &28.9            & 5.03            & 0.794 & 0.805& 11.163\\
 $^{50}$Fe($\beta^+){}^{50}$Mn  & $2_1^+, 0 $  & 7.499 & 23.03   & 3.81           & 0.976 & 0.868 & 3.096 \\
  $^{51}$Ca($\beta^-){}^{51}$Sc  & $3_1^-, 9 $  & 6.493 & 5.1   & 6.72           & 0.068 & 0.082 &14.522 \\
                                 & $1_1^-, 9 $  & 5.008 & 12.9  & 5.8          & 0.197 & 0.234 &   \\
                                  & $3_2^-, 9 $  & 4.646 & 9.9  & 5.77          & 0.204 & 0.297 &   \\   
  $^{51}$Sc($\beta^-){}^{51}$Ti  & $7_1^-, 7 $  & 3.819 & 2.03   & 6.180           & 0.548 & 0.585 &18.577 \\
  $^{51}$Ti($\beta^-){}^{51}$V  & $5_1^-, 5 $  & 2.152 & 91.9   & 4.90           & 0.556 & 0.542 &11.585 \\
                                & $3_1^-, 5 $  & 1.544 & 8.1   & 5.538           & 0.267 & 0.326 &   \\
 $^{51}$Cr($\beta^+){}^{51}$V  & $7_1^-, 5 $  & 0.753 & 90.06   & 5.391           & 0.447 & 0.487 &10.725 \\
 $^{51}$Mn($\beta^+){}^{51}$Cr & $7_1^-, 3 $  & 3.208 & 96.86  & 5.297           & 0.432 & 0.467 &5.363 \\
                               & $3_1^-, 3 $  & 2.459 & 0.2  & 7.186           & 0.049 & 0.467 &   \\
                               & $5_1^-, 3 $  & 1.855 & 0.015 & 7.303           & 0.043 & 0.055 &   \\
                               & $7_2^-, 3 $  & 1.651 & 0.01 & 7.07           & 0.056 & 0.060 &   \\
                               & $5_2^-, 3 $  & 1.207 &0.037 & 6.316           & 0.133 & 0.164 &   \\
                               & $7_3^-, 3 $  & 0.895 &0.09 & 5.662           & 0.283 & 0.311 &   \\
                                & $3_3^-, 3 $  & 0.379 &0.007 & 6.02           & 0.188 & 0.248 &   \\
 $^{51}$Fe($\beta^+){}^{51}$Mn & $7_1^-, 1 $  & 7.782 & 5.0  & 4.86           & 0.713 & 0.530 &5.363 \\
                               & $3_1^-, 1 $  & 6.194 & 0.49  & 5.32          & 0.420 & 0.327 &   \\
                               & $3_2^-, 1 $  & 5.879 & 0.24 & 5.51           & 0.338 & 0.465 &   \\
                               & $3_3^-, 1 $  & 5.105 & 0.10 & 5.54           & 0.326 & 0.532 &   \\
                               & $3_4^-, 1 $  & 4.464 &0.16 & 5.00          & 0.607 & 0.531 &   \\
$^{52}$Ca($\beta^-){}^{52}$Sc & $2_1^+, 10$ & 4.263 & 86.6           & 5.07           &0.229  & 0.437 & 7.584 \\ 
                                    & $2_2^+, 10$ & 1.634 & 1.4            & 5.80           &0.0987 & 0.198 &  \\ 
 $^{54}$Ca($\beta^-){}^{54}$Sc & $2^+, 12$   & 8.573 & 97(3)          & 4.25(2)        &0.588  & 0.683 & 8.192 \\ 
\hline
 \end{tabular}
   \end{center}
     \end{table*}

\addtocounter{table}{-1}

\begin{table*}
  \begin{center}
    \leavevmode
    \caption{{\em Continuation.\/}}   

    \label{tab:mgt_comp} 
   \begin{tabular}{lccccrrr}
    \hline
        &  & $Q$  & $I_\beta +
      I_\epsilon$ &  & \multicolumn{2}{c}{$M(GT)$} &  \\
      \cline{6-7}
     Process &$2J_n^\pi, 2T_n^\pi$ & (MeV) & (\%) & $\log ft$ & \multicolumn{1}{c}{Expt.} &
      \multicolumn{1}{c}{Theor.} & $W$ \\  
      \hline
      $^{54}$Sc($\beta^-){}^{54}$Ti & $4_1^+, 10$ & 10.504& 21(7)          & 5.7(2)         & 0.293 & 0.224 & 20.065 \\                                    
      $^{54}$Sc($\beta^-){}^{54}$Ti & $8^+, 10$   & 9.502 & 33(4)          & 5.3(1)         & 0.464 & 0.420 & \\ 
      $^{55}$Sc($\beta^+){}^{55}$Ti & $5^-,11$    &12.1   & 39(6)          & 5.0(2)         & 0.701 & 1.398 & 22.326 \\ 
                                    & $7^-,11$    & 11.691& 22(63          & 5.0(2)         & 0.701 & 1.198 &         \\
                                    & $9^-,11$    & 9954  & 11(1)          & 5.3(2)         & 0.496 & 0.827 &           \\
      $^{56}$V($\beta^-){}^{56}$Cr  & $0_1^+, 8$  & 9.2   &70              & 4.62           & 0.665 & 0.696 & 11.991  \\ 
                                    & $0_2^+, 8$  &  7.525& 26             & 4.63           & 0.657 & 0.834 &        \\
                                    & $4_1^+, 8$  & 8.194 &$<$4            & 5.63           & 0.208 & 0.190 &       \\
                                    & $4_2^+, 8$  &  7.37 & 1.0            & 6.01           & 0.134 & 0.304 &       \\
                                    & $4_3^+, 8$  &  6.876& 1.0            & 5.87           & 0.158 & 0.243 &       \\
     $^{58}$V($\beta^-){}^{58}$Cr   & $0^+, 10$   & 11.63 &$<$38           & 5.3            & 0.304 & 0.683 & 13.136  \\
                                    & $4^+, 10$   &  10.75&$<$34           & 5.3            & 0.304 & 0.834 &         \\
     $^{60}$Cr($\beta^-){}^{60}$Mn  & $2_1^+, 10$ & 6.46  & 88.6           & 4.2            & 0.623 & 1.136 & 7.584 \\
                                    & $2_2^+, 10$ & 5.701 & 10.2           & 5.0            & 0.078 & 0.369 &       \\
     $^{60}$Mn($\beta^-){}^{60}$Fe  & $0_1^+, 8$  & 8.444 & 88(2)          & 4.46(4)        & 0.780 & 0.928 & 11.991 \\
                                    & $0_2^+, 8$  & 6.47  & 5.0(6)         & 5.15(7)        & 0.361 & 0.756 &      \\
                                    & $0_3^+, 8$  & 6.088 & 3.0(5)         & 5.3(1)         & 0.304 & 0.336 &   \\
                                    & $4^+, 8$    & 7.621 &4.2(12)         & 5.6(2)         & 0.215 & 0.353 &   \\
     $^{61}$Cr($\beta^-){}^{61}$Mn  & $7^-, 11$   & 9.133 & 9(2)           & 5.6(1)         & 0.304 & 0.556 & 19.335 \\                                    
     $^{61}$Mn($\beta^-){}^{61}$Fe  & $3_1^-, 9$  &7.178  & 33(1)          & 5.02(3)        & 0.593 & 1.191 & 17.786 \\
                                    & $3_2^-, 9$  & 6.549 &39.0(6)         & 4.77(1)        & 0.791 & 1.285 &      \\
                                    & $3_3^-, 9$  & 5.925 &0.57(6)         & 6.40(5)        & 0.121 & 0.853 &        \\
                                    & $5_1^-, 9$  & 6.971 &12.6(8)         & 5.38(3)        & 0.392 & 0.698 &        \\
                                    & $5_2^-, 9$  & 6.017 &3.25(9)         & 5.64(1)        & 0.291 & 0.352 &        \\
                                    & $7^-, 9$    & 6.218 &0.49(7)         & 6.6(1)         & 0.096 & 0.112 &        \\
    $^{62}$Cr($\beta^-){}^{62}$Mn   & $2_1^+, 12$ & 7.77  &$\sim$72   & $\sim$4.2 & 0.623 & 0.801 & 8.192 \\
                                    & $2_2^+, 12$ & 7.13  &$\sim$25   & $\sim$4.4 & 0.495 & 0.954 &      \\
    $^{62}$Mn($\beta^-){}^{62}$Fe   & $8^+, 10 $  & 8.521 &8.4             & 5.9            & 0.264 & 0.242 & 22.752 \\
                                    & $6^+, 10 $  & 8.005 & 17             & 5.5            & 0.418 & 1.367 &         \\
                                   \hline
    \end{tabular}
  \end{center}
\end{table*}

The Gamow-Teller strength  B(GT) is calculated using following expression,
\begin{equation}
 {B(GT_{\pm})} = \frac{1}{2J_i + 1} f_q^2 \, |{\langle {f}|| \sum_{k}{\sigma^k\tau_{\pm}^k} ||i \rangle}|^2,
\end{equation}
where  $\tau_+|p\rangle = |n\rangle$ , $\tau_-|n\rangle  = |p\rangle$, the index $k$ runs over the single particle orbitals, 
$|i \rangle$ and $|f \rangle$ describe the state of the parent and daughter nuclei, respectively.
In the present work the B(GT) values
are scaled employing a quenching factor as we calculated from following formalism.

Following Ref. \cite{brownrgt,Mart} we define
\begin{equation}
M(GT)= [(2j_{i}+1)B(GT)]^{1/2},
\end{equation} 
this is independent of the direction of the transitions. In Table I we 
compared calculated and experimental $M(GT)$ values.
To get $R(GT)$, we need the total strength, W and it is defined by

\begin{equation}
  W=\left\{
  \begin{array}{@{}ll@{}}
    |g_{A}/g_{V}|[(2J_{i}+1)3|N_{i}-Z_{i}|]^{1/2} , & for N_{i} \neq  Z_{i},\\
    |g_{A}/g_{V}|[(2J_{f}+1)3|N_{f}-Z_{f}|]^{1/2} , & for N_{i} = Z_{i}. 
  \end{array}\right.
\end{equation} 

Where $R(GT)$ values define as  
\begin{equation}
R(GT) = M(GT)/W.
\end{equation} 

while the corresponding experimental versus the theoretical values are plotted in the Fig. 1.
The slope of this figure gives average quenching factor $q$ =0.768 $\pm$ 0.005. In the present work
we have performed calculations in $fp$ shell with two different effective interactions. Since the
overall results of KB3G effective interaction is better than GXPF1a, thus we have calculated
quenching factor only with KB3G effective interaction.

\section{Comparison of experimental and theoretical GT strength distributions}

In this section the theoretical results are compared with the experimental data reported in \cite{Molina} and
\cite{Adachi}.

\subsection{\bf$^{42}$Ti $\rightarrow$ $^{42}$Sc }

 Fig. ~\ref{42Ti} displays a comparison between the shell-model calculations and the experimental GT strength
 distribution for the transition $^{42}$Ti $\rightarrow$ $^{42}$Sc. 
Fig.~2(a) presents the experimental data observed through the
$\beta$-decay $^{42}$Ti$\rightarrow ^{42}$Sc up to the excitation energy $E_x(^{42}$Sc) = 1.888 MeV \cite{Molina}.
Fig.~2(b) shows the experimental data obtained
through the charge-exchange reaction $^{42}$Ca($^3${He},$t$)$^{42}$Sc up to the excitation
energy $E_x(^{42}$Sc) = 3.688 MeV \cite{Adachi}. Fig.~2(c) depicts the shell-model 
calculation using the KB3G interaction, Fig.~2(d), the shell-model calculation using the GXPF1a interaction,
and Fig.~2(e), the running sums of B(GT) as a function of the excitation energy. 

The experimental GT strength is dominated by the transition $^{42}$Ti$(0^+)$ $\rightarrow$ $^{42}$Sc(${1_1}^+$).
The reported energy $E_{1^+}$ is 611 keV, while the calculated ones are lower. The calculated intensities for
this transition are similar to the measured ones. It is noticeably that the interaction  KB3G generated an
excitation energy closer to the experimental one than the energy obtained employing the GXPF1a interaction,
while the opposite is true for the GT strength. The second excited $1^+$ state at 1888 keV is missed in both
calculations, which predict a second, small B(GT) strength at an excitation energy slightly above 4 MeV, which
could be the one observed in the CE reaction. Both interactions predict a noticeable B(GT) strength at an
excitation energy between 9 and 10 MeV, where there is no experimental information. The close similitude in
the B(GT) strength predicted using the GXPF1a interaction and the $\beta^+$ data is visible in the summed strength plot.

\begin{figure}
\begin{center}
\resizebox{0.50\textwidth}{!}{
\includegraphics{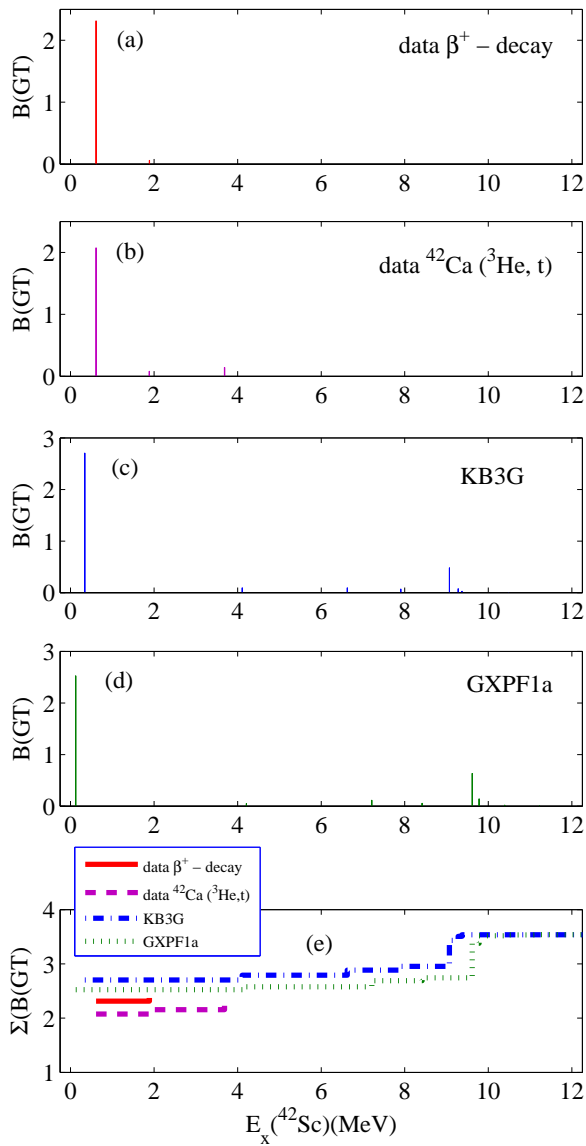} 
}
\end{center}
\caption{
Comparison of experimental and theoretical $B(GT)$ distributions for $^{42}$Ti.}
\label{42Ti}
\end{figure}

\subsection{\bf{$^{46}$Cr} $\rightarrow$ \bf{$^{46}$V}}

 \begin{figure}
\begin{center}
\resizebox{0.50\textwidth}{!}{
\includegraphics{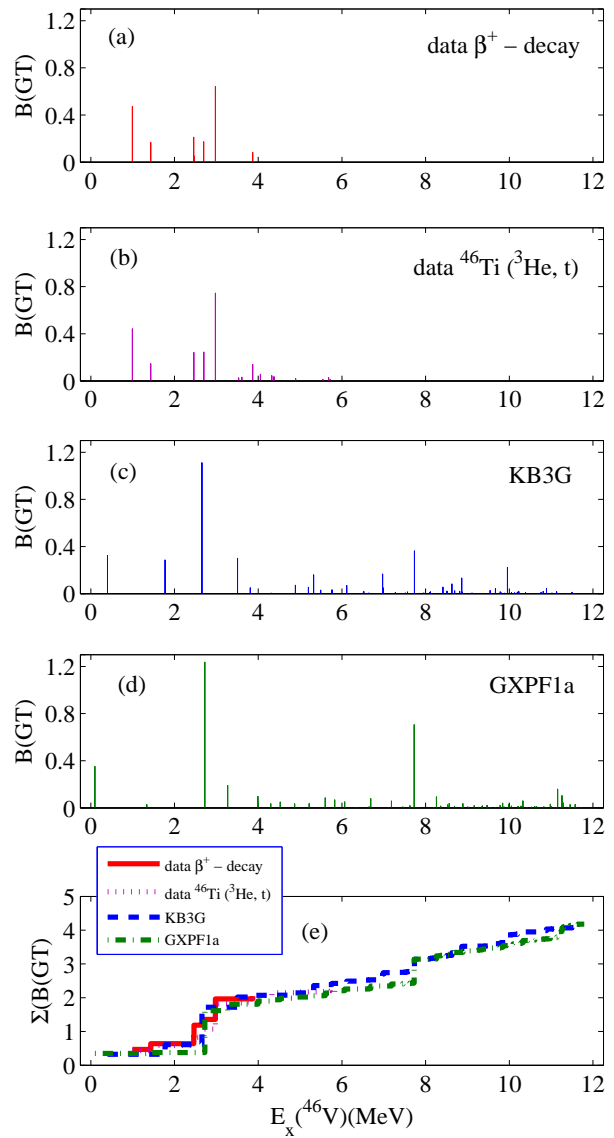} 
}
\end{center}
\caption{
Comparison of experimental and theoretical $B(GT)$ distributions for $^{46}$Cr.}
\label{46cr}
\end{figure}

Fig.~\ref{46cr} shows the experimental and shell-model calculated B(GT) strength distributions for the
transition $^{46}$Cr $\rightarrow$ $^{46}$V.
Fig.~3(a) represents the experimental data observed through the
$\beta$-decay $^{46}$Cr($0^+$) $\rightarrow$ $^{46}$V($1^+$) up to the excitation energy
$E_x(^{46}$V) = 3.867 MeV \cite{Molina}, Fig.~3(b) the experimental data observed through the
charge-exchange reaction process \cite{Adachi} i.e., $^{46}$Ti($^3${He},$t$)$^{46}$V up to the excitation
energy $E_x(^{46}$V) = 5.717 MeV, Fig.~3(c), the shell-model 
calculation using the KB3G interaction, Fig.~3(d), the shell-model calculation using the GXPF1a interaction,
and Fig.~3(e), the running sums of B(GT) as function of
excitation energy. 

The experimentally observed B(GT) strength as a function of the excitation energy exhibits two clusters,
one between 1 and 1.5 MeV, and another between 2.4 and 3.0 MeV, plus some small intensities around and above 4 MeV.
On the theoretical side, the KB3G and GXPF1a interactions predict a low energy transitions below 1 MeV, and the
most intense transition close to 3 MeV. While the general distribution of B(GT) strength is similar using both
interactions, the KB3G predicts more fragmentation. The summed B(GT) intensities obtained from the two calculations
are in close agreement, and reproduce well the observed one.

\subsection{\bf$^{50}$Fe $\rightarrow$ $^{50}$Mn }

\begin{figure}
\begin{center}
\resizebox{0.52\textwidth}{!}{
\includegraphics{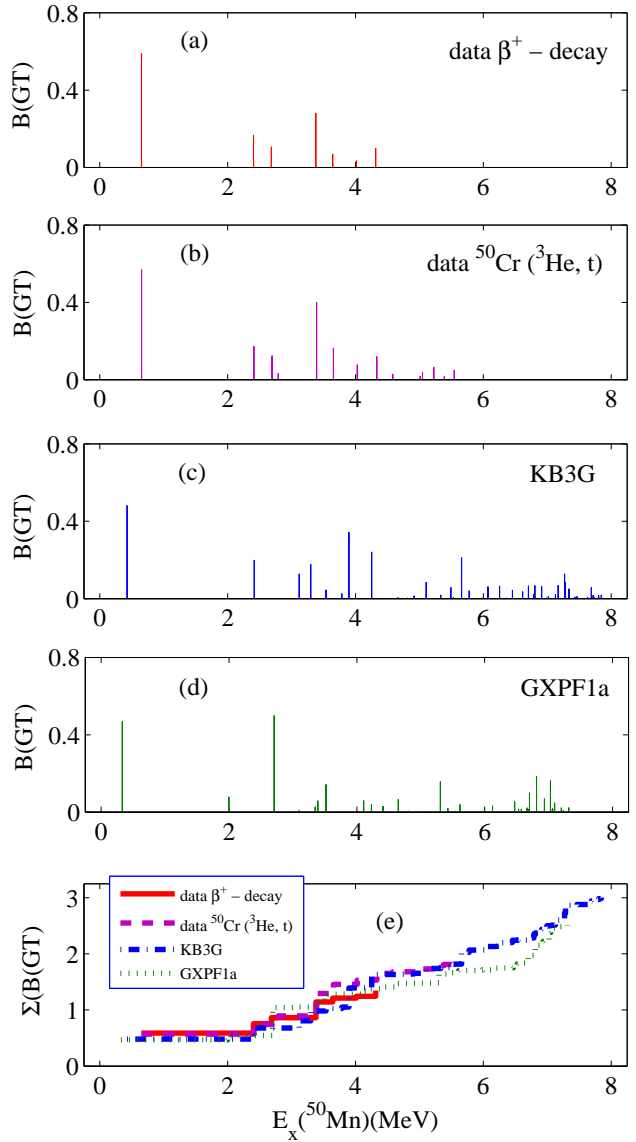} 
}
\end{center}
\caption{
Comparison of experimental and theoretical $B(GT)$ distributions for $^{50}$Fe.}
\label{50fe}
\end{figure}

The shell-model calculations and the experimental GT strength distributions for the transition
$^{50}$Fe $\rightarrow$ $^{50}$Mn are presented in the Fig.~\ref{50fe}.
The experimental data observed through 
the $\beta$-decay $^{50}$Fe$\rightarrow ^{50}$Mn up to the excitation energy $E_x(^{50}$Mn) = 4.315 MeV  \cite{Molina}
are shown in  Fig.~4(a), those observed through the  charge-exchange reaction process $^{50}$Cr($^3${He},$t$)$^{50}$Mn
up to the excitation energy $E_x(^{50}$Mn) = 5.545 MeV  \cite{Adachi} in Fig.~4(b),  the shell-model 
calculation using the KB3G interaction in Fig.~4(c), the shell-model calculation using the GXPF1a interaction in
Fig.~4(d), and the running sums of B(GT) as function of the excitation energy in Fig.~4(e).

There is an intense isolated B(GT) transition to the first $1^+$ state, observed at 651 keV, which is predicted,
but at lower excitation energies, by both interactions. There are a few observed transitions with comparable
strength distributed between 2.4 and 4.4 MeV, which are described with some detail using the interaction KB3G.
The same strength is concentrated in three transitions when using the interaction GXPF1a. Both interactions predict
a long tail of small intensity transitions. The calculated summed B(GT) intensities closely reproduce the experimental ones.  
  
\subsection{\bf$^{54}$Ni$\rightarrow^{54}$Co }

\begin{figure}
\begin{center}
\resizebox{0.50\textwidth}{!}{
\includegraphics{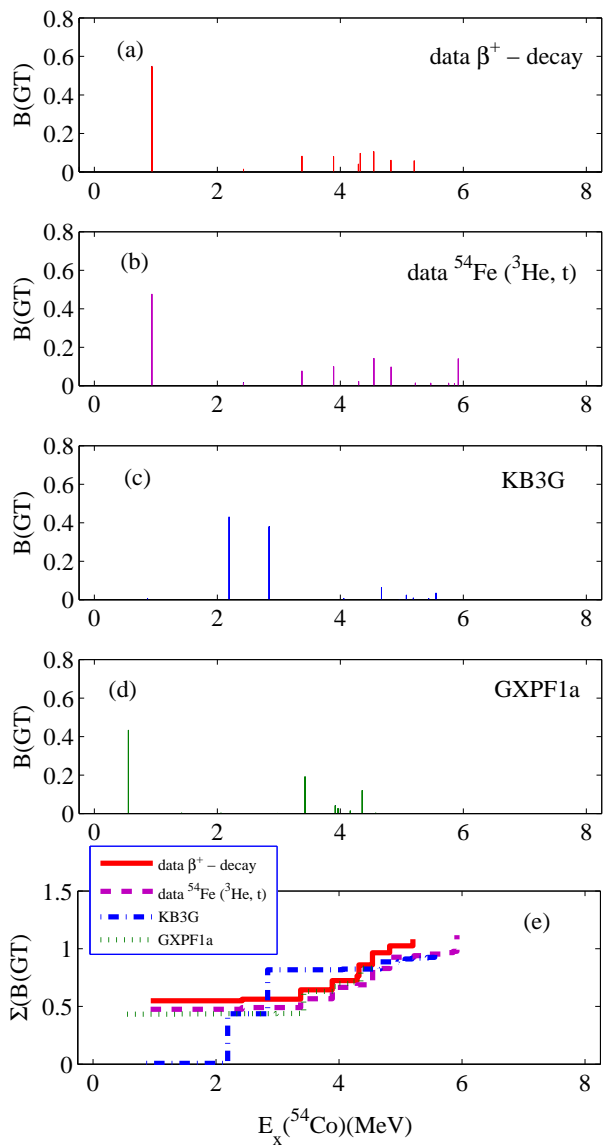} 
}
\end{center}
\caption{
Comparison of experimental and theoretical $B(GT)$ distributions for $^{54}$Ni.}
\label{54co}
\end{figure}

Fig.~\ref{54co} shows the experimental and shell-model calculated B(GT) strength distributions for the transition $^{54}$Ni$\rightarrow^{54}$Co.  
Fig.~5(a) displays the experimental data obtained through the
$\beta$-decay $^{54}$Ni$ \rightarrow ^{54}$Co up to the excitation energy $E_x(^{54}$Co) = 5.202 MeV  \cite{Molina},
Fig.~5(b) the experimental data observed through the charge-exchange reactions $^{54}$Fe($^3${He},$t$)$^{54}$Cr
up to the excitation energy $E_x(^{54}$Co) = 5.917 MeV  \cite{Adachi}, Fig.~5(c) the shell-model calculation using
the KB3G interaction, Fig.~5(d), the shell-model calculation using the GXPF1a interaction, and Fig.~5(e),
the running sums of  B(GT) as function of the excitation energy. 

The B(GT) strength for the $^{54}$Ni($0^+$) $\rightarrow$ $^{54}$Co($1_1^+$) transition displays a dominant
transition at 937 keV and a set of transitions at energies between 3.3 and 6 MeV.
As mentioned above, in the shell  model calculations a truncation to a maximum of four nucleon excitations
from the $f_{7/2}$ shell to the rest of the $pf$ orbitals was necessary due to computational limitations.
The B(GT) strength distribution obtained employing the KB3G interaction in the truncated space fails to
reproduce the experimental data. On the other hand, the calculated B(GT) obtained with the GXPF1a interaction
depict the main elements observed in the experiments. The intense low energy transition is present, although
at a slightly lower energy, and two transitions around 4 MeV resemble the centroid of the observed ones. 

The sum of B(GT) strength naturally follows that same pattern. The results from the KB3G interaction do not
resemble the observed distribution, while those associated to the GXPF1a interaction are in good agreement
with experimental data even with truncated calculation.
Due to the huge matrix dimensions we calculated only ten transitions from ground state of  $^{54}$Ni($0^+$)
to $^{54}$Co($1^+$).

\section{The summed B(GT) strength }

In Table~\ref{ESPM}, the total sum of the B(GT) strength is presented for the  transitions measured in the
four nuclei using the shell model code ANTOINE \cite{antoine}.  The third and fourth columns show the measured valued for the $\beta$-decay and the charge exchange
($^3${He},$t$) reactions, respectively. Both experimental results are of the same order,
their differences can be ascribed to the different energy regions accessible with these techniques. The last
three columns show the calculated
results obtained employing the KB3G interaction, the GXPF1a interaction and the extreme single particle model (ESPM), respectively.

\begin{table}[h]
\caption{\label{tab:table1}Comparison between the experimental, SM calculation, and ESPM summed B(GT) strengths.
Here we have reported unquenched summed $B(GT)$ values for shell model.}
\label{ESPM}
\begin{center}
\begin{tabular}{|c|c||c|c||c|c|c|}
\hline  & & \multicolumn{4}{c}{$\sum_i B(GT+)_i$}&\\
\hline 
  $ $ & $(Z, N)$ & $\beta$-decay & CER & KB3G & GXPF1a & ESPM \\ \hline
  $^{54}$Ni & (28, 26) & 1.082 & 1.117  & 12.197 & 13.362  & 16.29 \\
  $^{50}$Fe & (26, 24) & 1.344 & 1.859 & 9.464  & 10.277  & 14.14 \\
  $^{46}$Cr & (24, 22) & 2.047& 2.219 & 7.231   & 7.613 & 10.70 \\
  $^{42}$Ti & (22, 20) &  2.372 & 2.297  & 6.000   &  6.000  & 6.00 \\
\hline
\end{tabular}
\end{center}
\end{table}

In the extreme single particle model (ESPM) the $0^+$ ground state of the even-even parent nuclei is described filling the
$f_{7/2}$ orbital with the appropriate number of valence protons and neutrons.
The final $1^+$ states in the odd-odd daughter nuclei
are built as a hole in the proton $f_{7/2}$ shell, and a neutron particle in any of the $pf$ orbitals.

 The Gamow-Teller strengths are calculated in the ESPM as
\begin{equation}
  B(GT+)_{i} = \frac{1}{3} {n_p} {n_i} {|\langle {f_{7/2}}|{\sigma\tau}_+ |i \rangle|}^2
\end{equation}
In this expression ${n_p}$ is the number of  valence protons in  the $f_{7/2}$ shell, 
$n_i$ the number of valence neutron holes in the $i-th$ orbital, which in this case can only be the
$f_{7/2}$ (non-spin flip transition) and  the $f_{5/2}$ (spin flip transition). 
${|\langle {f_{7/2}}|{\sigma\tau}_+ |i \rangle|}^2$ is single-particle matrix element connecting
the proton state $f_{7/2}$ and the neutron state $i$.

It is clear from the table that the extreme single particle summed B(GT) strengths are much larger
than the observed ones. Those obtained in the SM calculations are closer to the experimental intensities.
The comparison between ESPM and shell model results for each transition suggesting that we
need separate quenching factor for each concerned nuclei.

\section{Conclusions }

In the present work we have presented a comprehensive shell model calculation for Gamow-Teller transition
strengths in  $^{42}$Ti, $^{46}$Cr, $^{50}$Fe and $^{54}$Ni, employing the effective interactions KB3G and GXPF1a.
They provide a theoretical description of the experimental
Gamow-Teller transition strength distributions measured via $\beta$ decay of these $T_z$=-1 nuclei,
produced in fragmentation at GSI, and also with ($^3${He},$t$) charge-exchange (CE) reaction.

In the study of the GT transitions in $^{42}$Ti, $^{46}$Cr, $^{50}$Fe, the configuration space of the
full $pf$ shell was employed. Both interactions provided a qualitative description of the observed transitions,
and were able to closely reproduce  the summed B(GT) strength.

In the case of $^{54}$Ni  it was necessary to impose a truncation in the number of excitations allowed from
the $f_{7/2}$ level. 
The theoretical strengths are larger than the experimental value. It may mean that a substantial amount of strength has 
 not been experimentally measured.

In all cases the calculations predict the existence of a fragmented but observable B(GT) strength at excitation
energies between 6 to 12 MeV, which could become observable in future experiments. 
Present work will also add more information to ref. \cite{Fujita2014}.

\vspace{2mm}

VK acknowledge financial support from CSIR-INDIA for his PhD thesis work. We thank Prof. Jorge G. Hirsch
for his support during this work. We also thank Prof. Y. Fujita for his comments on this article.


\begin{thebibliography}{99}
\bibitem{Fujita2011}
Y. Fujita {\it et al.}, Prog. Part. Nucl. Phys. {\bf 66}, 549 (2011), and references therein.
\bibitem{Rolfs88} C.E. Rolfs, W. Rodney, {\em Cauldrons in the Cosmos}
University of Chicago Press (1988).
\bibitem{Langanke2003}
K. Langanke and G. Mart\'inez-Pinedo, Rev. Mod. Phys. {\bf 75}, 819 (2003).
%
\bibitem{Balasi2015} K.G. Balasi, K. Langanke, G. Mart\'inez-Pinedo, Progress in Particle and Nuclear Physics
{\bf 85}, 33 (2015)
%
%
\bibitem{Fujita2005}
Y. Fujita {\it et al.}, Phys. Rev. Lett. {\bf 95}, 212501 (2005).
\bibitem{Fujita2014} Y. Fujita {\it et al.}, Phys. Rev. Lett. {\bf 112}, 112502 (2014).

\bibitem{Orrigo2014}
S.E.A. Orrigo {\it et al.}, Phys. Rev. Lett. {\bf 112}, 222501 (2014).
\bibitem{LZ1} M. Harper and L. Zamick,  Phys. Rev. C {\bf 91}, 014304 (2015).
\bibitem{LZ2} M. Harper and L. Zamick,  Phys. Rev. C {\bf 91}, 054310 (2015).
\bibitem{talmi} I. Talmi, Adv. Nucl. Phys. {\bf 27}, 1 (2003).
\bibitem{martinez1} E. Caurier, G. Mart\'inez-Pinedo, F. Nowacki, A. Poves, and A.P. Zuker, Rev. Mod. Phys. {\bf 77}, 427 (2005).
\bibitem{richtler} A. Richtler, Nucl. Phys. A {\bf 507}, 99 (1990).
\bibitem{kota}V.K.B. Kota, R. Sahu, K. Kar, J. M. G. G\'omez, and J. Retamosa, Phys. Rev. C {\bf 60}, 051306(R) (1999).
\bibitem{caurier} E. Caurier, A. Poves, and A. P. Zuker, Phys. Rev. Lett. {\bf 74}, 1517 (1995).
\bibitem{mar} T. Marketin, G. Mart\'inez-Pinedo, N. Paar, and D. Vretenar, Phys. Rev. C {\bf 85}, 054313 (2012).
\bibitem{sjq} S.J.Q. Robinson, Ph.D. Thesis, Rutgers University (2002).
\bibitem{zamick1} S.J.Q. Robinson and L. Zamick, Phys. Rev. C {\bf 66}, 034303 (2002).
\bibitem{zamick2} R. Garcia and L. Zamick, Phys. Rev. C {\bf 92}, 034322 (2015).
\bibitem{pai} H. Pai {\it et al.}, Phys. Rev. C {\bf 93}, 014318 (2016).
\bibitem{petermann} I. Petermann, G. Mart\'inez-Pinedo, K. Langanke, and E. Caurier, Eur. Phys. J. A {\bf 34}, 319 (2007).
\bibitem{heyde} K. Heyde, P. von Neumann Cosel and A. Richter, Rev. Mod. Phys. {\bf 82}, 2365 (2010).
\bibitem{ber} G.F. Bertsch and I. Hamamoto, Phys. Rev. C {\bf 26}, 1323 (1982).
\bibitem{nat} C.B. Hinke {\it et al.}, Nature {\bf 486}, 341 (2012).
\bibitem{Caurier99}
E. Caurier, K. Langanke, G. Mart\'inez-Pinedo, and F. Nowacki, Nucl. Phys. A {\bf 653}, 439 (1999).
\bibitem{Suzuki11}
T. Suzuki {\it et al.}, Phys. Rev. C {\bf 83}, 044619 (2011).
\bibitem{Cole2012}
A.L. Cole {\it et al.}, Phys. Rev. C {\bf 86}, 015809 (2012).
\bibitem{Molina} F. Molina {\it et al.}, Phys. Rev. C {\bf 91}, 014301 (2015).
\bibitem{Molinathesis} F. G. Molina Ph.D Thesis "Beta decay of $T_z = -1$ nuclei and comparison with charge exchange
reaction experiments", (2011).
\bibitem{Adachi} T. Adachi {\it et al.}, Phys. Rev. C {\bf 73}, 024311 (2006).
\bibitem{Honma2} M. Honma  {\it et al.}, Eur. Phys. J. A {\bf 25}, (s01) 499 (2005).
\bibitem{Brown}  B. A. Brown, W. D. M. Rae, E. McDonald, and M. Horoi, NushellX@MSU.
\bibitem{Honma1} M. Honma {\it et al.}, Phys. Rev. C {\bf 65}, 061301(R) (2002).
\bibitem{Honma3} D.-C. Dinca et al., Phys. Rev. C {\bf 71}, 041302(R) (2005).
\bibitem{Honma4} B. Fornal et al., Phys. Rev. C {\bf 70}, 064304 (2004).
\bibitem{A.Poves} A. Poves {\it et al.}, Nucl. Phys. A {\bf 694}, 157 (2001).

\bibitem{brownrgt}
B.A. Brown and B.H. Wildenthal, At. Data Nucl. Data Tables \textbf{33}, 347 (1985).
\bibitem{Mart} G. Martinez-Pinedo {\it et al.}, Phys. Rev. C {\bf 53}, R2602(R) (1996).
\bibitem{nndc} Electronic version of Nuclear Data Sheets, http://www.nndc.bnl.gov/.
\bibitem{antoine} E. Caurier,  G. Mart\'inez-Pinedo , F. Nowacki, A. Poves,
and A. P. Zuker, Rev.\ Mod.\ Phys. {\bf77} (2005) 427.






\end{thebibliography}
\end{document}